\documentclass[prl,superscriptaddress,twocolumn,showpacs,preprintnumbers,amsmath,floatfix]{revtex4}
\usepackage{amssymb}
\usepackage{graphicx}

\begin{document}

\title{Revision of model parameters for $\kappa$-type charge transfer salts:
 an {\it ab initio} study.}

\author{Hem C. Kandpal}
\affiliation{Institut f\"ur Theoretische Physik, Goethe-Universit\"at
Frankfurt, Max-von-Laue-Stra{\ss}e 1, 60438 Frankfurt am Main, Germany}
\author{Ingo Opahle}
\affiliation{Institut f\"ur Theoretische Physik, Goethe-Universit\"at
Frankfurt, Max-von-Laue-Stra{\ss}e 1, 60438 Frankfurt am Main, Germany}
\author{Yu-Zhong Zhang}
\affiliation{Institut f\"ur Theoretische Physik, Goethe-Universit\"at
Frankfurt, Max-von-Laue-Stra{\ss}e 1, 60438 Frankfurt am Main, Germany}
\author{Harald O. Jeschke}
\email{jeschke@itp.uni-frankfurt.de}
\affiliation{Institut f\"ur Theoretische Physik, Goethe-Universit\"at
Frankfurt, Max-von-Laue-Stra{\ss}e 1, 60438 Frankfurt am Main, Germany}
\author{Roser Valent{\'\i}}
\affiliation{Institut f\"ur Theoretische Physik, Goethe-Universit\"at
Frankfurt, Max-von-Laue-Stra{\ss}e 1, 60438 Frankfurt am Main, Germany}

\date{\today}

\begin{abstract}
  Intense experimental and theoretical studies have demonstrated that the
  anisotropic triangular lattice as realized in the
  $\kappa$-(BEDT-TTF)$_2$X family of organic charge transfer (CT) salts
  yields a complex phase diagram with magnetic, superconducting, Mott
  insulating and even spin liquid phases. With extensive density
  functional theory (DFT) calculations we refresh the link between manybody
  theory and experiment by determining hopping parameters of the
  underlying Hubbard model. This leads us to revise the widely used
  semiempirical parameters in the direction of less frustrated,
  more anisotropic
  triangular lattices. The implications of these results on the
  systems'  description are discussed.
\end{abstract}


\maketitle

A strong research trend of the new millennium has been the desire to
understand complex manybody phenomena like superconductivity and
magnetism by realistic modelling, {\it i.e.} to employ precise first
principles calculations to feed the intricate details of real
materials into the parameter sets of model Hamiltonians that are then
solved with increasingly powerful manybody techniques.  The
$\kappa$-(BEDT-TTF)$_2$X~\cite{ET} organic charge transfer salts are a
perfect example for a class of materials with such fascinating
properties that they drive progress in experimental and manybody
methods alike.  Experimentally, the phase diagram shows Mott
insulating, superconducting, magnetic and spin liquid
phases~\cite{Elsinger00,Shimizu03,Kurosaki05,Kagawa05}.
Theoretically, the underlying anisotropic triangular lattice is a
great challenge due to effects of frustration and the intense efforts
to get a grip on the problem include studies with path integral
renormalization group (PIRG)~\cite{Morita02}, exact
diagonalization~\cite{Clay08}, variational Monte
Carlo~\cite{Watanabe06}, cluster dynamical mean field
theory~\cite{Kyung06,Ohashi08} and dual Fermions~\cite{Lee08} to cite a few.  In this rapidly
expanding field of research, electronic structure calculations play
the decisive role of mediating between the complex underlying
structure and phenomenology of organic charge transfer salts and the
models used for understanding the physics~\cite{Powell06}, and in this
work, we will provide the perspective of precise, state of the art
electronic structure calculations.

Previously, $\kappa$-type CT salts have been investigated by
semi-empiricial and first principles electronic structure
calculations. The most commonly used $t$, $t'$, $U$ parameter sets
derive from extended H\"uckel 
 molecular orbitals
calculations~\cite{Komatsu96,Fortunelli97} performed on different
constellations of BEDT-TTF dimers.

The main result reported here is that our  first
principles study shows all four considered $\kappa$-type CT salts 
 to be {\it less}
frustrated than  previously assumed  based on semiempirical theory. Most
importantly, the often cited value of $t'/t=1.06$~\cite{Komatsu96} for
the  spin liquid material $\kappa$-(ET)$_2$Cu$_2$(CN)$_3$ should be replaced by the
significantly smaller value $t'/t=0.83\pm 0.08$.  This has
fundamental implications on the systems' model description 
 as we shall see below.

In this Letter, we employ the Car-Parrinello~\cite{Car85}
projector-augmented wave~\cite{Bloechl94} molecular dynamics
(CPMD) method for relaxing the only partially known structures and
perform structure optimizations at constant
pressure~\cite{Parrinello80} in order to prepare high pressure
structures. We calculated 
the
electronic structure using two
full potential all-electron codes, the  linearized
augmented plane wave (LAPW) method implemented in WIEN2K~\cite{wien2k}
and the 
full potential 
local-orbital (FPLO) method~\cite{Koepernik99}.

We base our study on the following structures: For
$\kappa$-(BEDT-TTF)$_2$Cu$_2$(CN)$_3$, the crystal
structure~\cite{Geiser91a} was published without hydrogen atom
positions; we add them and perform a CPMD relaxation within the
generalized gradient approximation (GGA) in which we keep lattice
parameters and heavy atom (S, Cu) positions fixed~\cite{PAWpar}. With this procedure
we make sure that the important atomic positions in the crystal
structure remain as given by the diffraction measurements.  We remove
the inversion center in the middle of a cyano group by lowering the
symmetry from $P2_1/c$ to $Pc$. No high pressure structures of
$\kappa$-(BEDT-TTF)$_2$Cu$_2$(CN)$_3$ are published, and we therefore
perform a full relaxation of the structure and lattice parameters at a
pressure $P = 0.75$~GPa~\cite{pressure}; we recover the $Pc$ symmetry by further
optimizing the high pressure structure with symmetry constraints. For
$\kappa$-(BEDT-TTF)$_2$Cu(SCN)$_2$, structures at ambient pressure and
at $P=0.75$~GPa are known~\cite{Rahal97}. Ambient pressure structures
are available for $\kappa$-(BEDT-TTF)$_2$Cu[N(CN)$_2$]Br~\cite{Geiser91b,Kini90}
and $\kappa$-(BEDT-TTF)$_2$Cu[N(CN)$_2$]Cl~\cite{Williams90}. 
 We will
refer to the four materials as $\kappa$-CN, $\kappa$-SCN, $\kappa$-Br
and $\kappa$-Cl for brevity.  At ambient pressure and low temperatures
$\kappa$-SCN and $\kappa$-Br superconduct while $\kappa$-Cl orders
antiferromagnetically.

\begin{figure}[tb]
\includegraphics[width=0.48\textwidth]{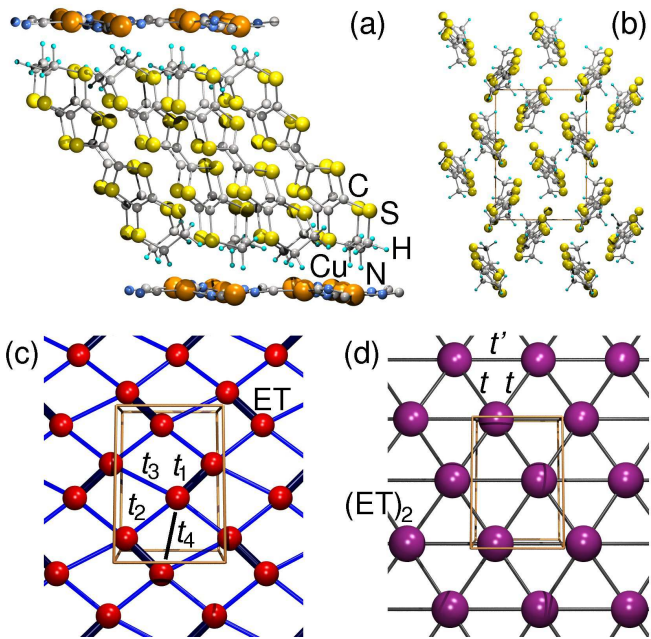}
\caption{(Color online) Structure of
  $\kappa$-(BEDT-TTF)$_2$Cu$_2$(CN)$_3$ as typical example of the
  $\kappa$-type CT salts. (a) Side view with ET layers separated by
  insulating Cu$_2$(CN)$_3$ anion layers. (b) ET lattice, viewed in the $bc$
  plane.
 The unit cell containing four ET molecules is
  shown. (c)  ET lattice shown in the same projection as (b)
where the ET molecules have been replaced by single spheres. ET
  dimers are emphasized by heavier bonds. (d) Lattice of ET dimers,
  replaced by single spheres. }
\label{fig:str}
\end{figure}

As an example for the structure of all $\kappa$-type CT salts
considered here, we show in Fig.~\ref{fig:str} the structure of
$\kappa$-CN. The strongly anisotropic nature of the materials is
apparent from the alternation of charge donating ET layers and
acceptor anion layers (Fig.~\ref{fig:str}~(a)), Cu$_2$(CN)$_3$ in this
case. Experimental observations~\cite{Komatsu96} as well as our
calculations show that Cu occurs in Cu$^{+1}$ oxidation state here and
has a filled $3d$ shell, making the anion layers insulating. The
$\kappa$-type arrangement of ET molecules (Fig.~\ref{fig:str}~(b))
exhibits a strong dimerization of the molecules, four of which are
contained in a unit cell. They form the lattice shown in simplified
form in (Fig.~\ref{fig:str}~(c)). In the charge transfer process, each
molecule donates a charge of $-0.5 e$ and is left with half a hole.
Considering pairs of dimers as the fundamental unit leads to the
triangular lattice of Fig.~\ref{fig:str}~(d) with one hole per dimer.

\begin{figure}[tb]
\includegraphics[angle=-90,width=0.48\textwidth]{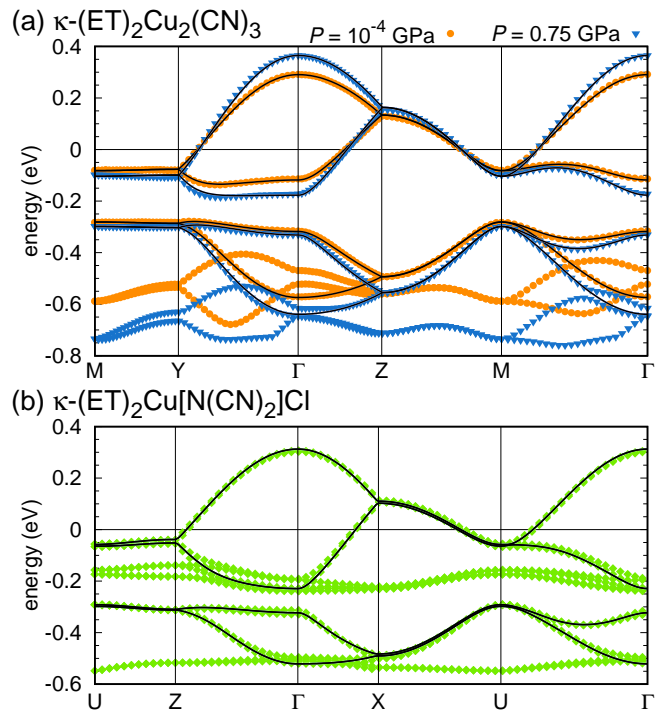}
\caption{(Color online) DFT bandstructures of (a)
  $\kappa$-(BEDT-TTF)$_2$Cu$_2$(CN)$_3$ at ambient pressure and at
  $P=0.75$~GPa, and of (b) $\kappa$-(BEDT-TTF)$_2$Cu[N(CN)$_2$]Cl
  shown with symbols. Lines represent the tight binding fits to the
  bands close to $E_{\rm F}$ deriving from ET molecules.  }
\label{fig:bs}
\end{figure}

In Fig.~\ref{fig:bs} we show three examples of low energy
bandstructures for $\kappa$-CN at ambient and elevated pressure
(Fig.~\ref{fig:bs}~(a)) and for $\kappa$-Cl (Fig.~\ref{fig:bs}~(b)).
The overall features of the bandstructures are very similar to the
known semiempirical band structures, with two antibonding bands of the
ET molecules crossing the Fermi level, and the corresponding bonding
bands nearly mirror symmetric below. Due to the double number of eight
ET molecules per unit cell in the case of $\kappa$-Cl,
Fig.~\ref{fig:bs}~(b) actually has four bands crossing the Fermi
level, which are pairwise degenerate due to the high symmetry ($Pnma$)
of the crystal structure.  In contrast to previous calculations, there
are four bands with mainly Cu $3d$ character close to the ET bands;
slightly below the bonding ET bands in the case of $\kappa$-CN and
slightly below the antibonding ET bands in the case of $\kappa$-Cl.
 The
bandstructures of Fig.~\ref{fig:bs}~(a) were performed with LAPW using
the GGA functional~\cite{LAPWpar}. We checked the reliability and reproducibility of
these results by repeating the calculation with FPLO~\cite{FPLOpar}, and we found
that the band structures and all derived quantities match very well.
As $\kappa$-Cl has roughly twice as many atoms in the unit cell
compared to $\kappa$-CN, we used the faster FPLO method, within GGA,
for Fig.~\ref{fig:bs}~(b). The low energy bands show almost no
dispersion in the direction perpendicular to the anion layers which is
the $a$ direction (M-Y in the Brillouin zone) for $\kappa$-CN and the
$b$ direction (U-Z in the Brillouin zone) for $\kappa$-Cl, confirming
the quasi-two-dimensional character of the $\kappa$-type CT salts.

\begin{figure}[tb]
\includegraphics[width=0.45\textwidth]{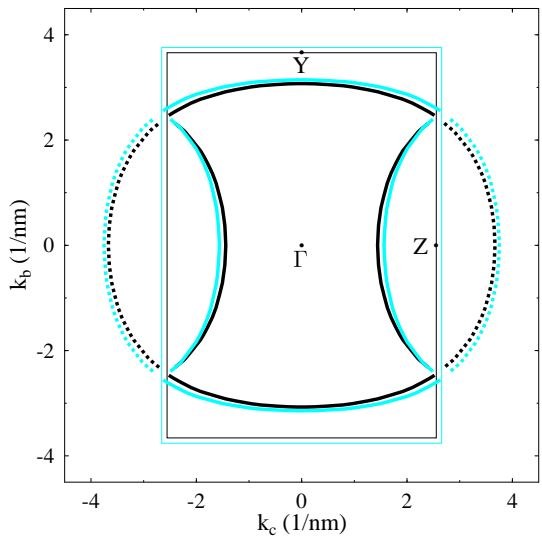}
\caption{(Color online) Fermi surface cuts for nonmagnetic
$\kappa$-(BEDT-TTF)$_2$Cu$_2$(CN)$_3$ at ambient pressure (black)
and at $P=0.75$~GPa (light). Thin lines show the respective first 
Brillouin zone. The combined elliptical orbit as detected by AMRO 
measurements (see text) is indicated by broken lines.
}
\label{fig:fscut}
\end{figure}
We corroborated that the calculated bandstructures agree with
experimental evidence by calculating the Fermi surface (FS) for
$\kappa$-CN at ambient pressure and at $P=0.75$~GPa. The FS consists
of two different sheets arising from the two bands crossing the Fermi
energy, a corrugated cylinder around $Z$ and a quasi one dimensional
sheet parallel to $M-Y$. Both sheets touch along the line $M-Z$ and
give rise to a combined elliptical orbit (Fig.~\ref{fig:fscut}) which
encloses 99-100\% of the area of the first Brillouin zone.  This is in
good agreement with previous calculations~\cite{Komatsu96} and with
angle-dependent magnetoresistance oscillations (AMRO)
measurements~\cite{Ohmichi97}.  Our calculated FS shows the same
behavior with pressure as observed in experiment.  The area enclosed
by the combined elliptical orbit increases by 6\% from 37.4~nm$^{-2}$ at
ambient pressure to 39.7~nm$^{-2}$ at $P=7.5$~kbar (see
Fig.~\ref{fig:fscut}), which is consistent with the
increase of 6\% between 2.1 and 7~kbar obtained from AMRO
measurements in Ref.~\cite{Ohmichi97}.


In the next step, we use the DFT band structures to extract the
parameters of the Hubbard Hamiltonian
\begin{equation}\begin{split}
    H &= \sum_{<ij>,\sigma} t (c_{i\sigma}^\dag
    c_{j\sigma}^{\phantom{\dag}}
    +{\rm H.c.})+\sum_{[ij],\sigma} t' (c_{i\sigma}^\dag c_{j\sigma}^{\phantom{\dag}}+{\rm H.c.})\nonumber\\
    &+U\sum_i \Big(n_{i\uparrow} -\frac{1}{2}\Big)\Big(n_{i\downarrow}
    -\frac{1}{2}\Big)\,.
\label{eq:H}
\end{split}\end{equation}
where $<ij>$ and $[ij]$ indicate sums over nearest and next nearest
neighbors, respectively.  We do this by fitting the bands to a tight
binding model. Note that also nonlocal correlations $V$, $V'$ are
thought to play a role in organic CT salts but their 
determination 
is beyond the scope of our present investigation. We consider each
molecule position as shown in Fig.~\ref{fig:str}~(c) as a site; then
the four (eight) sites per unit cell contribute the four (eight)
ET bands at the Fermi
level in the case of $\kappa$-CN and $\kappa$-SCN  ($\kappa$-Br and
$\kappa$-Cl)
 This yields the hopping integrals
$t_1$ to $t_4$ where the index indicates increasing distance; $t_1$
is the intradimer hopping integral. The
 black lines in Fig.~\ref{fig:bs} show as an example the tight-binding
fit obtained from the molecular model.  We follow
Ref.~\onlinecite{Komatsu96} in the use of the geometrical formulas
$t=(t_2+t_4)/2$ and $t'=t_3/2$ for obtaining $t$ and $t'$ of
Eq.~\ref{eq:H}. In order to corroborate these results, we use the
alternative method of considering the ET dimers as sites of the tight
binding model as shown in Fig.~\ref{fig:str}~(d). Then the two dimers
per unit cell for $\kappa$-CN and $\kappa$-SCN are responsible for the two
antibonding ET bands at the Fermi level (or four dimers/four
antibonding bands in the case of $\kappa$-Br and $\kappa$-Cl).

\begin{figure}[tb]
\includegraphics[angle=-90,width=0.48\textwidth]{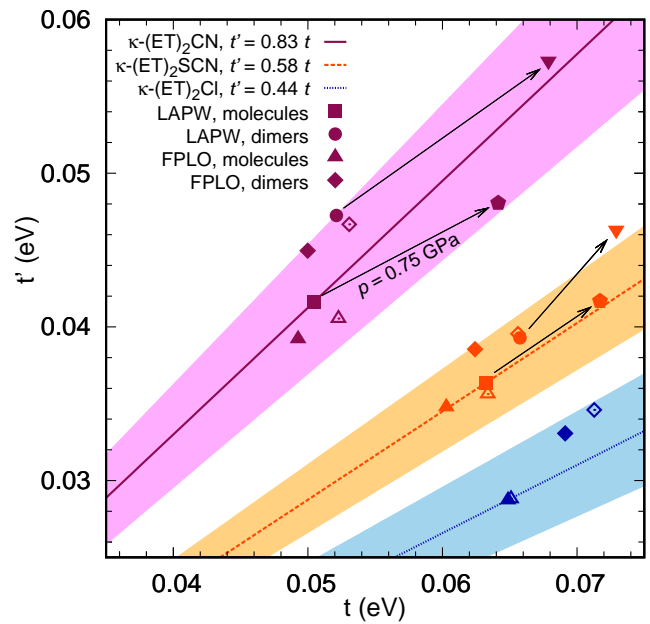}
\caption{(Color online) Overview of hopping parameters obtained from
  GGA (filled symbols) and LDA (open symbols) calculations for
  $\kappa$-(BEDT-TTF)$_2$Cu$_2$(CN)$_3$,
  $\kappa$-(BEDT-TTF)$_2$Cu(SCN)$_2$ and
  $\kappa$-(BEDT-TTF)$_2$Cu[N(CN)$_2$]Cl. The spread of $t$ and $t'$
  parameters depending on basis set, functional and number of fitted
  bands yields $t'/t$ ratios with a margin of error indicated by lines
  and shaded regions. Arrows indicate the change of $t'/t$ ratio due
  to application of a pressure of $P=0.75$~GPa.}
\label{fig:ttp}
\end{figure}

The result of these extensive bandstructure calculations and tight
binding fits is summarized in Fig.~\ref{fig:ttp}. The choice of basis
set (LAPW or FPLO), functional (GGA or LDA) and number of sites
included in the fit (molecules or dimers) results in a certain spread
of resulting $t$, $t'$ pairs which we interpret as a margin of error.
 The slopes of the
lines connecting the origin with the tight binding results
for the  GGA bands~\cite{comment}
show the results for the $t'/t$ ratios at ambient pressure: $t'/t=0.83\pm
0.08$ for $\kappa$-CN, $t'/t=0.58\pm 0.05$ for $\kappa$-SCN,
$t'/t=0.44\pm 0.05$ for $\kappa$-Cl, and $t'/t=0.42\pm 0.08$ for
$\kappa$-Br.

With the tight binding results, we can estimate the Hubbard $U$ as
$U\approx 2t_1$ following Ref.~\cite{McKenzie98}.  We obtain $U/t$
ratios of $U/t=7.3$ for $\kappa$-CN, $U/t=6.0$ for $\kappa$-SCN,
$U/t=5.5$ for $\kappa$-Cl and $U/t=5.1$ for $\kappa$-Br. These values
provide a readjustment of the position of the four $\kappa$-type CT
salts in the $(U/t,t'/t)$ phase diagram, compared to the semiempirical
calculations: For $\kappa$-CN, we replace $(8.2,1.06)$ by
$(7.3,0.83)$, for $\kappa$-SCN $(6.8,0.84)$ by $(6.0,0.58)$, for
$\kappa$-Cl $(7.5,0.75)$ by $(5.5,0.44)$ and for $\kappa$-Br
$(7.2,0.68)$ by $(5.1,0.42)$. At $P=0.75$~GPa, we have $(6.0,0.75)$ for
$\kappa$-CN and $(5.7,0.58)$ for $\kappa$-SCN.

In Fig. ~\ref{fig:ttp} we also show the analysis of pressure effects
on the $t$, $t'$ parameters.  We represented with arrows the parameter
changes from the ambient pressure results to the $P=0.75$~GPa results
in $\kappa$-SCN and $\kappa$-CN.  For $\kappa$-CN, pressure leads to a
decrease in the $t'/t$ ratio, while we observe almost constant $t'/t$
for $\kappa$-SCN.  On the other hand we can establish that the
parameter differences among the three studied cases at ambient
pressure can be attributed to internal chemical pressure due to the
different anion sizes. By comparing the ratios $t'/t$ for $\kappa$-CN,
$\kappa$-SCN and $\kappa$-Cl at ambient pressure with the physical
pressure $t'/t$ behavior for $\kappa$-CN and $\kappa$-SCN we observe
that chemical pressure and physical pressure have the same effect on
the $t'/t$ ratio: for $\kappa$-CN, pressure significantly decreases
the $t'/t$ ratio. This effect is weaker for $\kappa$-SCN.

We now consider the new positioning of the four $\kappa$-type CT salts
in some of the calculated $(U/t,t'/t)$ phase diagrams. In the PIRG
phase diagram of Ref.~\onlinecite{Morita02}, $\kappa$-Cl are shifted
from the nonmagnetic insulating (NMI) phase to the antiferromagnetic
insulating (AFI) phase which is an improvement as the superconducting
phase is not considered in this work. Meanwhile, $\kappa$-CN and
$\kappa$-SCN are only repositioned within the NMI phase. In the CDMFT
phase diagram of Ref.~\onlinecite{Kyung06}, $\kappa$-Br is moved from
the AF phase to the d-wave superconducting (SC) phase; $\kappa$-Cl is
moved from the spin liquid (SL) phase to the border between AF and
d-wave superconducting (SC) phase. $\kappa$-CN is moved from the
d-wave SC phase to the border between SL and SC phases, and
$\kappa$-SCN is moved within the SC phase. The effect of pressure
means that $\kappa$-CN is moved diagonally across the SC phase by the
application of $P=0.75$~GPa. In particular, the positions of
$\kappa$-Br, $\kappa$-Cl and $\kappa$-CN in the phase diagram are
clearly improved.  Our results exclude the possibility of viewing
$\kappa$-CN as a quasi-one-dimensional system~\cite{Hayashi07}.

In conclusion, we have presented the results of density functional
theory calculations for four $\kappa$-type charge transfer salts. We
obtain important shifts of the familiar $t'/t$ ratios from
semiempirical electronic structure calculations~\cite{Komatsu96}
towards significantly smaller values, {\it i.e.} towards lower
frustration. Due to the smaller overall band width of the ET bands at
the Fermi level, our estimate for the $U/t$ values is also below the
prevalent values. Our results call for a reexamination of the
description of the $\kappa$-type charge transfer salts, and in
particular for $\kappa$-CN, namely which is the nature of the spin
liquid at ambient pressure as well as the phase transition from Mott
insulator to superconducting state under pressure.


{\em Note added:} While finalizing this manuscript, Nakamura {\em et
al.}~\cite{Nakamura09} have posted a manuscript, where similar values
of $t'/t$ for $\kappa$-CN and $\kappa$-SCN as in our full-potential
calculations are obtained from a pseudopotential method.

We acknowledge useful discussions with L. Tocchio and C. Gros.  We thank the
Deutsche Forschungsgemeinschaft for financial support through the
TRR/SFB~49 and Emmy Noether programs and we acknowledge support by the
Frankfurt Center for Scientific Computing.


\begin{thebibliography}{99}

\bibitem{ET} 
  BEDT-TTF, or shorter ET, stands for
  bis(ethylene-dithio)tetrathiafulvalene.

\bibitem{Elsinger00}
  H. Elsinger {\it et al.}, Phys. Rev. Lett. {\bf 84}, 6098 (2000).

\bibitem{Shimizu03} 
  Y. Shimizu {\it et al.}, Phys. Rev. Lett. {\bf 91}, 107001 (2003).

\bibitem{Kurosaki05}
  Y. Kurosaki {\it et al.}, Phys. Rev. Lett. {\bf 95}, 177001 (2005).

\bibitem{Kagawa05}
  F. Kagawa {\it et al.}, Nature {\bf 436}, 534 (2005).

\bibitem{Morita02}
  H. Morita {\it et al.}, J. Phys. Soc. Jpn. {\bf 71}, 2109 (2002).

\bibitem{Clay08}
  R. T. Clay {\it et al.}, Phys. Rev. Lett. {\bf 101}, 166403 (2008).

\bibitem{Watanabe06} 
  T. Watanabe {\it et al.}, J. Phys. Soc. Jpn. {\bf 75}, 074707
  (2006), and Phys. Rev. B {\bf 77}, 214505 (2008).

\bibitem{Kyung06}
  B. Kyung, A.-M. S. Tremblay, Phys. Rev. Lett. {\bf 97}, 046402
  (2006).

\bibitem{Ohashi08}
  T. Ohashi {\it et al.}, Phys. Rev. Lett. {\bf 100}, 076402 (2008).

\bibitem{Lee08}
  H. Lee {\it et al.}, Phys. Rev. B {\bf 78}, 205117 (2008).

\bibitem{Powell06}
  B. J. Powell, R. H. McKenzie, J. Phys.: Condens. Matter {\bf 18},
  R827 (2006).

\bibitem{Komatsu96}
  T. Komatsu {\it et al.}, J. Phys. Soc. Jpn. {\bf 65}, 1340 (1996).

\bibitem{Fortunelli97}
  A. Fortunelli, A. Painelli, J. Chem. Phys. {\bf 106}, 8051 (1997),
  and Phys. Rev. B {\bf 55}, 16088 (1997).


\bibitem{Car85} 
  R. Car, M. Parrinello, Phys. Rev. Lett. {\bf 55}, 2471 (1985).

\bibitem{Bloechl94} 
  P. E. Bl{\"o}chl, Phys. Rev. B {\bf 50}, 17953 (1994).

\bibitem{Parrinello80} 
  M. Parrinello, A. Rahman, Phys. Rev. Lett. {\bf 45}, 1196 (1980).

\bibitem{wien2k} P. Blaha {\it et al.} WIEN2K, An Augmented Plane
  Wave+Local Orbitals Program for Calculating Crystal Properties
  (Karlheinz Schwarz/Techn. Universit\"at Wien, Wien, Austria, 2001).

\bibitem{Koepernik99} 
  K. Koepernik, H. Eschrig, Phys. Rev. B {\bf 59}, 1743 (1999);
  http://www.FPLO.de

\bibitem{Geiser91a}
  U. Geiser {\it et al.}, Inorg. Chem {\bf 30}, 2586 (1991).

\bibitem{PAWpar} 
  For CP-PAW, we employ a $(4\times 4\times 4)$ $k$ mesh and plane
  wave cutoffs of 60~Ryd and 240~Ryd for the wavefunction and charge
  density, respectively.

\bibitem{pressure} 
  In the constant pressure~\cite{Parrinello80} CPMD method, the
  pressure $P$ directly enters the Lagrangian for the lattice dynamics
  as a term $P\Omega$ with unit cell volume $\Omega$.

\bibitem{Rahal97}
M. Rahal {\it et al.}, Acta Cryst. B {\bf 53}, 159 (1997).

\bibitem{Geiser91b}
  U. Geiser {\it et al.}, Acta Cryst. C {\bf 47}, 190 (1991).

\bibitem{Kini90} 
  A. M. Kini {\it et al.}, Inorg. Chem. {\bf 29}, 2555 (1990).

\bibitem{Williams90}
  J. M. Williams {\it et al.}, Inorg. Chem. {\bf 30}, 3272 (1990).

\bibitem{LAPWpar}
  For LAPW, we employ a $(5\times 10\times 7)$ $k$ mesh in the
  irreducible wedge of the Brillouin zone, integrated by tetrahedron
  method~\cite{Bloechl94b} and a value $R_{MT} \times k_{\rm max}
  =3.37$ due to the presence of hydrogen.

\bibitem{Bloechl94b} 
  P. E. Bl\"ochl {\it et al.}, 
  Phys.  Rev. B {\bf 49}, 16223 (1994).

\bibitem{FPLOpar}
  FPLO convergency was tested with up to 350 $k$ points in the full
  Brillouin zone, using the tetrahedron method~\cite{Bloechl94b}.

\bibitem{Ohmichi97}
  E. Ohmichi {\it et al.}, J. Phys. Soc. Jpn. {\bf 66}, 310 (1997).

\bibitem{comment} The lines are connecting the origin with the
results corresponding to the
tight-binding fit  to four LAPW GGA (eight FPLO GGA) 
bands for $\kappa$-CN and $\kappa$-SCN 
($\kappa$-Cl), and the shaded regions containing
all other results represent the possible deviation.

\bibitem{McKenzie98}
  R. H. McKenzie, Comments Condens. Matter Phys. {\bf 18}, 309 (1998).

\bibitem{Hayashi07}
  Y. Hayashi, M. Ogata, J. Phys. Soc. Jpn. {\bf 76}, 053705 (2007).

\bibitem{Nakamura09} K. Nakamura {\it et al.}, arXiv:0903.5409.
\end{thebibliography}
\end{document}